\documentclass[aps,prb,showpacs,twocolumn]{revtex4}
\usepackage{graphicx}  % needed for figures
\usepackage{dcolumn}   % needed for some tables
\usepackage{bm}        % for math
\usepackage{amssymb}   % for math
\usepackage{t1enc}     % Danish letters

\begin{document}
\title{First-principles Analysis of Photo-current in Graphene PN Junctions}
\date{\today}
\author{Jingzhe Chen}
\email{jingzhe.chen@gmail.com}\affiliation{Centre for the Physics of
Materials and Department of Physics, McGill University, Montreal,
PQ, Canada, H3A 2T8}

\author{Yibin Hu}
\affiliation{Centre for the Physics of Materials and
Department of Physics, McGill University, Montreal, PQ, Canada, H3A 2T8}

\author{Hong Guo}
\affiliation{Centre for the Physics of Materials and
Department of Physics, McGill University, Montreal, PQ, Canada, H3A 2T8}

\begin{abstract}
We report a first principles investigation of photocurrent generation by graphene PN junctions. The junctions are formed by either chemically doping with nitrogen and boron atoms, or by controlling gate voltages. Non-equilibrium Green's function (NEGF) formalism combined with density functional theory (DFT) is applied to calculate the photo-response function. The graphene PN junctions show a broad band photo-response including the terahertz range. The dependence of the response on the angle between the light polarization vector and the PN interface is determined. Its variation against photon energy $E_{ph}$ is calculated in the visible range. The essential properties of chemically doped and gate-controlled PN junctions are similar, but the former shows fingerprints of dopant distribution.
\end{abstract}

\maketitle

\section{Introduction}

The extraordinary electronic and optical properties of graphene make it a promising material for novel applications in photonics and optoelectronics. \cite{Nature_Photonics} Experimental investigations have discovered several very interesting behaviors of graphene - some are quite unique, such as the universal optical conductance, high transparency\cite{Optical_Conductance}, and broadband non-linear photoluminescence\cite{photoluminescence}. Possible applications include transparent electrodes, photovoltaic cells, and touch-screens\cite{Transparent_Electrode, Transparent_Electrode1}. One of the most interesting applications is the graphene photo-detector. The gapless nature of graphene implies that there may always be an electron-hole pair in resonance with any excitation, allowing a broad operational wavelength ranging from 300$nm$ to 6$\mu m$ which can overcome the long-wavelength limit of conventional semiconductors\cite{Terahertz, Photoconductivity_graphene}. Once excited, due to the high carrier mobility of graphene, hot electrons can transfer energy to the entire system in very short time scales ($\sim 1ps$)\cite{lifetime,relaxation}, and a 10$Gbit s^{-1}$ optical data link for error-free detection of optical data stream has been reported\cite{Communication,Fast_response}. These results demonstrate a great potential of graphene to be an advanced material for light sensors and high speed communications.

A dipole potential is necessary for the photovoltaic (PV) process that produces a net dc electric current. For a graphene system, it can be achieved by either tuning a gate voltage locally or chemically doping acceptor and donor atoms at different regions. Photocurrent produced by both schemes was reported  previously\cite{PNP_imaging1, PNP_imaging2, Dopped_exp}. The chemically doped graphene PN junction is more common in experimental studies; the gate controlled junction is more difficult to fabricate since it requires more precise setup of the device structure and light illumination. For the chemical junctions, the doping level lies within hundreds of millivolts which corresponds to terahertz or far-infrared frequencies; on the other hand, the gate controlled junction is able to reach the higher frequency regime.

The photo-response behavior of graphene PN junctions is very important and warrant systematical theoretical investigations. Recently, Refs.\cite{Model,ModelLong} reported theoretical analysis of gate controlled graphene PN junctions by solving a model Hamiltonian and determining photo-response within a semi-classical transport model. To the best of our knowledge, photo-response of graphene PN junctions have not been investigated from atomistic first principles point of view. It is the purpose of this work to fill this gap and we investigate the photo-response function for both the chemical and gate controlled systems. In particular, photo-response in a broad frequency range including the terahertz is investigated. Such a broad band photo-response is difficult to reach by conventional semiconductors. The dependence of photo-response on the angle between the direction of light polarization and the PN interface is determined. This dependence is found to be well consistent with the previous theoretical result\cite{Model}. We also determined the dependence of photo-response on photon energy $E_{ph}$ at the range of $E_{ph} \sim U$, where U is the potential decline in the depletion region of the PN junction. In the solar visible range, for $E_{ph} < U$ the photo-response is almost linear, agreeing with that found in Ref.\cite{ModelLong}, and it tends to saturate when $E_{ph}$ exceeds $U$. Signatures of dopant distribution can be found in the photo-response function of the chemical junctions. By investigating both chemical doping and gate control, the PV properties of graphene PN junctions are compared and a comprehensive understanding is achieved. Finally, our atomistic approach provides a benchmark result for first-principle optoelectronic studies of the graphene system.

The rest of the paper is organized as the following. In the next section, the calculation method is presented. Sections III and IV present results for the chemically doped and gate controlled junctions, respectively. The last section is reserved a for short discussion and summary.

\section{Method}

\begin{figure}[ht]
\centering
\includegraphics[width=0.5\textwidth]{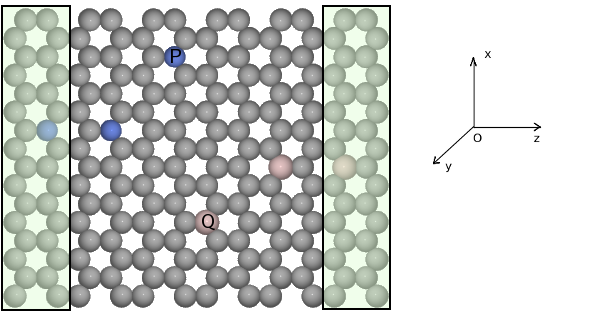}
%LCR2.png
\caption{(color online) Schematic atomic structure of the PN junction. The system is a graphene sheet doped with N (blue) atoms on the left side, and B (pink) atoms on the right. The regions in shadow are part of the electrodes which extend periodically to $z\pm \infty$. DC current flows from the left to the right (along the z-axis) or vice versa. In the first principles calculations, the device is treated as periodic in the x- and y-directions. In the y-direction, a large vacuum region is included in the calculation super-cell that effectively isolates interactions between the graphene sheet and its periodic images. \textbf{P} and \textbf{Q} represent positions of dopant atoms at the interface.}
\label{fig1}
\end{figure}

A two-probe graphene PV device consists of a scattering region sandwiched by the left and right graphene electrodes. Fig.\ref{fig1} plots such a two-probe PV system showing a chemically doped graphene PN junction. Here, a PN junction is formed by doping nitrogen (N, blue) on the left of the junction and boron (B, pink) on the right. The scattering region of the device consists of the PN junction and many layers of (doped) graphene on the left and right. Far from the PN junction, the scattering region connects to the left and right electrodes: the electrode atoms are shown in the shadowed boxes. The electrodes extend to electron reservoirs at $z=\pm \infty$ where photo-current are collected. In the analysis, the scattering region should be large enough (12 atomic layers in our calculations) so that the electronic structure of the electrodes are not affected by the charge transfer at the PN junction in the middle of the structure.

Because current flow is intrinsically a nonequilibrium transport problem, our calculation is based on carrying out real space density functional theory (DFT) calculation within the Keldysh nonequilibrium Green's functions (NEGF) formalism\cite{taylor}. Very briefly, in the NEGF-DFT formalism, DFT is used to self-consistently calculate the electronic structure and Hamiltonian of the device; NEGF is used to determined the quantum statistical information that is needed to populate the electronic structure and calculate the density matrix; real space numerical methods are used to handle the transport and electro-static boundary conditions at the interface between the electrodes and the device scattering region. Since the NEGF-DFT method has been well documented, we refer interested readers to the original literature for more details\cite{taylor}. In this work, a GPAW implementation\cite{GPAW} of the NEGF-DFT is used for the calculations.

Our interest in this work is to analyze the photo-response of the PV device. A linear polarized light is shined on the scattering region and not the electrodes. This is reasonable because very long graphene nanoribbon (> 1000nm) can be fabricated experimentally\cite{dai1} and the electrodes can be covered by optically nontransparent materials. To obtain the corresponding photocurrent, we first calculate the Hamiltonian $\hat{H}_0$ of the two-probe PV device (e.g. Fig.\ref{fig1}) without light, using the NEGF-DFT self-consistent method\cite{taylor,GPAW}. Here, the exchange-correlation is treated with the PBE functional\cite{PBE} and a (4,1) k-mesh is applied the in x-y plane for k-sampling. A single zeta polarized (SZP) atomic orbital basis set is used to expand all the physical quantities. For graphene, SZP basis is sufficient for obtaining accurate results.

After $\hat{H}_0$ is self-consistently calculated, electron-photon interaction is added to it as a perturbation for the subsequent analysis of photocurrent. This way, the total Hamiltonian of the electron-photon system is:
\begin{equation}
\hat{H}=\hat{H}_0 + \frac{e}{m_0}\mathbf{A}\cdot\hat{\mathbf{p}}
\end{equation}
where $A$ is the polarization vector of the light. Assuming the electromagnetic field of a photon is a single-mode monochromatic plane wave, also assuming the susceptibility and dielectric constants to be homogenous, we have \cite{Henri}
\begin{equation}
\mathbf{A}(t)=\hat{a}(\frac{\hbar\sqrt{\tilde{\mu_r}\tilde{\epsilon_r}}}
{2N\omega\tilde{\epsilon}c}I_{\omega})^{1/2}(be^{-i\omega t}
+b^{\dagger}e^{i\omega t})
\end{equation}
where $\mu_r$, $\epsilon_r$, $\epsilon$ are the relative susceptibility, relative dielectric constant and absolute dielectric constant, respectively. $\hat{a}$ represents polarization of the field, $I_\omega$ is the photon flux defined as the number of photons per unit time per unit area,
\begin{equation}
I_{\omega}\equiv\frac{Nc}{V\sqrt{\tilde{\mu_r}\tilde{\epsilon}_r}}\ \ .
\end{equation}

The first order electron-photon Keldysh self-energies are \cite{Henri,Datta},
\begin{equation}
\Sigma_{lm}^{\gtrless}(E)=\sum_{pq}M_{lp}M_{qm}[NG^{0\gtrless}_{pq}
(E\pm\hbar\omega)+(N+1)G^{0\gtrless}_{pq}(E\mp\hbar\omega)]
\label{eq4}
\end{equation}
where $M_{lm}\equiv\frac{e}{m_0}(\frac{\hbar\sqrt{\tilde{\mu_r}\tilde{\epsilon_r}}}
{2N\omega\tilde{\epsilon}c}I_\omega)^{1/2}<l|p_z|m>$,
assuming $z$ to be the light polarization direction. The Keldysh Green's
functions are written as:
\begin{equation}
G^<=G^{0r}(i\Gamma_Lf_L+i\Gamma_Rf_R+\Sigma_{ph}^<)G_{0a}\ ,
\label{G<}
\end{equation}
\begin{equation}
G^>=G^{0r}(i\Gamma_L(1-f)_L+i\Gamma_R(1-f)_R+\Sigma_{ph}^>)G_{0a}
\label{G>}
\end{equation}
where $\Gamma$ represents the coupling of the device scattering region to the electrodes, namely the linewidth function of the electrodes. $G^{0r}$ and $G^{0a}$ are the retarded and advanced Green's functions (without photons), respectively. In Eq.(\ref{eq4}), $G^{0\gtrless}$ is calculated by Eq.(\ref{G<},\ref{G>}) without the electron-photon interaction.

With the ground state Hamiltonian $\hat{H}_0$ of the two-probe PV device, quantities $\Gamma$, $G^{0r}$, $G^{0a}$, $G^{0\gtrless}$ can be calculated\cite{taylor}. Afterward, $\Sigma^{\gtrless}$ is calculated via Eq.(\ref{eq4}), which can then be put into Eq.(\ref{G<}) and Eq.(\ref{G>}) to determine the Keldysh Green's function $G^{\gtrless}$.

Finally, following Ref.\onlinecite{jauho}, the linear phase-coherent photocurrent is calculated by the following formula,
\begin{eqnarray}\nonumber
I_{ph}&=&\frac{ie}{2\hbar}\int\frac{d\epsilon}{2\pi}
Tr\left\{[\Gamma^L(\epsilon)-\Gamma^R(\epsilon)]G^<(\epsilon)+\right.\\
&&\left.[f_L(\epsilon)\Gamma^L(\epsilon)-f_R(\epsilon)
\Gamma^R(\epsilon)][G^r(\epsilon)-G^a(\epsilon)]\right\}\ .
\end{eqnarray}
Since the photon field is taken as a perturbation to $\hat{H}_0$, the photocurrent obtained this way is in the small intensity regime and nonlinear optical effects are not considered in this work.

\section{Chemically doped junctions}

For the chemically doped graphene PN junction, the schematic atomic structure is shown in Fig.\ref{fig1} where a hetero-junction is formed by doping the two sides with nitrogen or boron atoms, respectively. We investigated six junctions with different dopant configurations (inset in Fig.6). For these junctions, while it is more realistic to use metal leads far away from the PN junction to collect current and connect the PN junction to the outside world, to reduce the computational demand we extended the doped graphene to $z=\pm\infty$ to act as device leads. This is acceptable because photocurrent is generated in the PN junction where the dipole potential locates.

\begin{figure}[ht]
\centering
\includegraphics[width=0.5\textwidth]{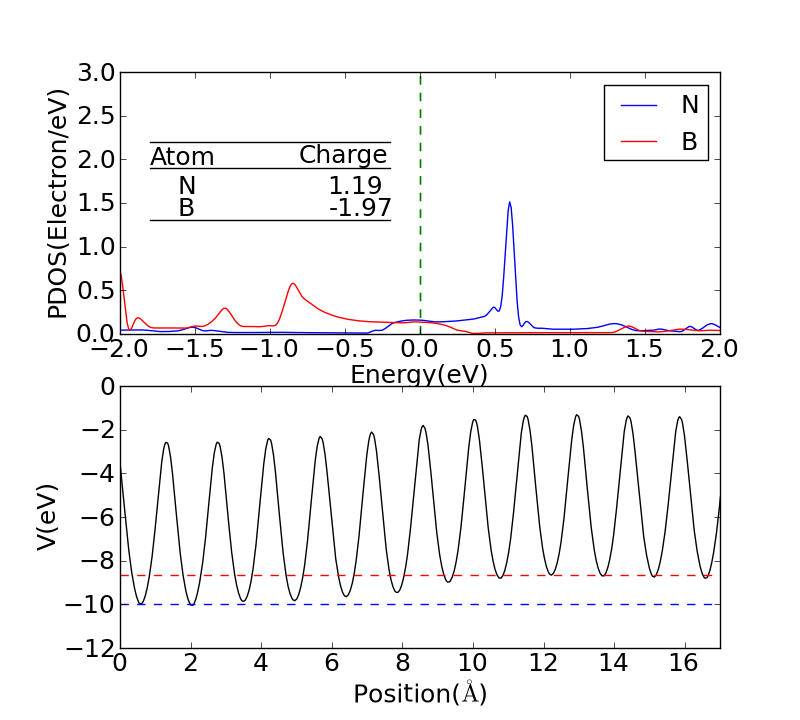}
%NB_pdos_vt.png
\caption{(color online) Upper panel: the PDOS of N and B atoms. There is a donor peak of N around 0.6eV (blue peak) and an acceptor peak of B around -0.9eV (red peak). The dashed vertical line at energy zero is the Fermi level. The result of Bader charge transfer analysis is listed in the inset table. Lower panel: averaged effective potential profile along the transport direction. Blue dashed line represents the level in the left electrode and red dash line for the right.}
\label{fig2}
\end{figure}

To be specific, let's consider a particular configuration (the sixth pattern in the inset of Fig.6). Fig.\ref{fig2} plots the calculated ground state properties without photons. The partial density of states (PDOS) in the upper panel of Fig.\ref{fig2} shows two characteristic peaks which are contributed by the donor (N) and acceptor (B) atoms just above and below the Fermi level (dashed vertical line). As a donor, the N atoms give electrons to the graphene which is indicated by its PDOS peak being above the Fermi level (blue peak near 0.6eV). Similarly, the B atoms accept electrons and its PDOS peak is located below the Fermi level (red peak near -0.9eV). On the other hand, a Bader charge analysis shows that the N atom actually gains some charge, 1.19e on average; and the B atom loses about 1.97e (see table in the upper panel of Fig.\ref{fig2}) - these are in agreement with previous literature\cite{charge_transfer}. Hence the Bader charge analysis appears to contradict to N and B atoms being donors and acceptors. A further detailed analysis showed that the carbon atoms in the N side of the junction lose electrons in total but gain electrons in their \textit{$p_y$} orbital which forms conjugate $\pi$ states for charge carriers that contribute to transport. On the B side of the junction, the carbon atoms lose $\pi$ electrons. In other words, the N atoms lose some high energy level charges as donor while attract more low energy level covalent electrons due to greater electron affinity. The situation of B atoms is analogous to that of N. This way, by separating the transport carriers (the $\pi$ electrons) from the total electron number on the atoms, the Bader charge analysis is reconciled with the roles of N and B atoms being donors and acceptors.

The carbon atoms in the system, to some extent, follow the behavior of the dopants, i.e. the C atoms in the N or B side form an extra conduction or valence band, respectively. Then, an electron in the extra valence bands (contributed by (B) and [C(B)]) can absorb a photon and hop to the extra conduction bands (contributed by (N) and [C(N)]) across the Fermi level, eventually giving rise to the electron flow from the B to the N side of the PN junction. This is also reflected in the average effective potential shown in the lower panel of Fig.\ref{fig2}, where we observe a potential decline about $\sim1.4eV$ at the interface (the oscillation is due to the ionic potential). An electron-hole pair is generated after absorbing a photon, the electron flows along the potential decline and the hole in the opposite direction. A photocurrent is therefore generated in the graphene PN junction by the photovoltaic effect without external bias.

\begin{figure}[ht]
\centering
\includegraphics[width=0.5\textwidth]{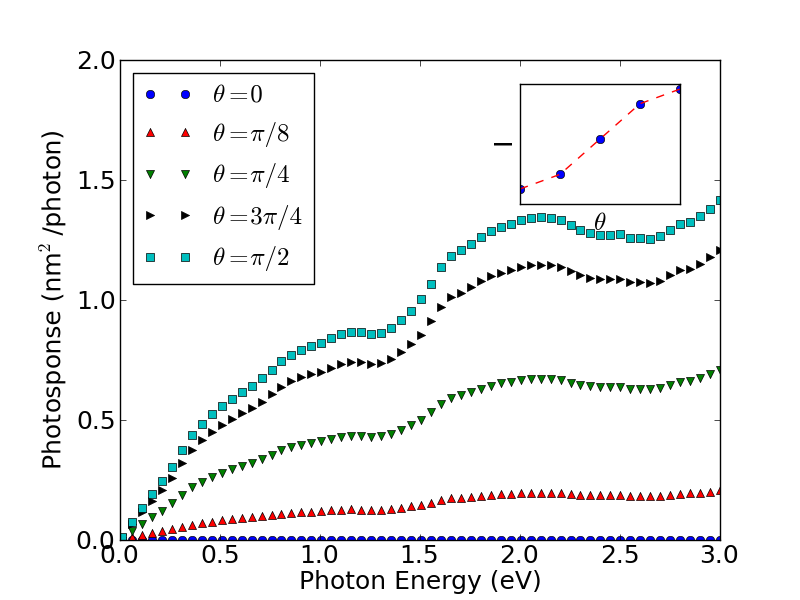}
%Directions.png}
\caption{(color online) Photo-response function with different light polarization directions. The polarization direction changes continuously from the $\pi$ orbital direction (y-axis) to the transport direction (z-axis). Inset: photo-current at photon energy 2.0eV, versus different polarization angles. The red dashed line is fitting to the trigonometric function $I_{\pi/2}sin^2\theta$.}
\label{fig3}
\end{figure}

We now turn on the light and investigate the dependence of photo-response to light polarization. Fig.\ref{fig3} plots the photo-response function,
\begin{equation}
f\equiv \frac{I_{ph}}{eF}
\label{f1}
\end{equation}
versus photon energy with different polarization directions of the light. Here $F$ is the photon flux: in our perturbation theory, photocurrent scales with $F$.  $\theta$ in Fig.\ref{fig3} is the angle between the direction of light polarization and the y-axis. The photocurrent almost vanishes completely when $\theta=0$, i.e., when the electric field of the light vibrates along the y axis. This is because the angular momentum quantum number of such a photon is equal to one along the y-axis, while both the valence band in the B side and the conduction band in the N side of the PN junction are in the $\pi$ orbital, also having unity angular momentum quantum number along the y-axis. The photon induced hoping from the valence band to the conduction band is therefore prohibited due to the mismatch of the three angular quantum numbers. This limitation starts to be relieved when $\theta$ increases from zero and is completely removed when $\theta$ reaches $\pi/2$ at which the maximum photocurrent is reached. As a direct result of the Fermi's golden rule, the photocurrent scales with $sin^2\theta$ (inset of Fig.\ref{fig3}) which is in good agreement with results of Ref.\onlinecite{Model}. Since the angle dependence is determined by the angular momentum of the $\pi$ orbital, the gate controlled PN junctions to be discussed below show the same trend.

Different from the discrete features predicted for carbon nanotubes\cite{Leonard_tube}, photo-response of graphene is smooth over a broad frequency range corresponding to the continuous spectrum of the graphene material (e.g. Dirac bands near the Fermi level). In other words, there is no obvious transition features between discrete states in the $f-E_{ph}$ curve. Photons with energy $E_{ph}$ can excite electrons in the range [$-E_{ph}$,$0$] to generate electron-hole pairs; these pairs can be separated in the depletion region of the PN junction and contribute to the photocurrent.

We have also investigated the role of dopant distribution by comparing results of six samples having different dopant locations (only the positions indicated by letters P and Q in Fig.\ref{fig1} are randomly changed). These results will be compared with those of the gate controlled PN junctions in the next section.

\section{Gate controlled junction}

\begin{figure}[ht]
\centering
\includegraphics[width=0.45\textwidth]{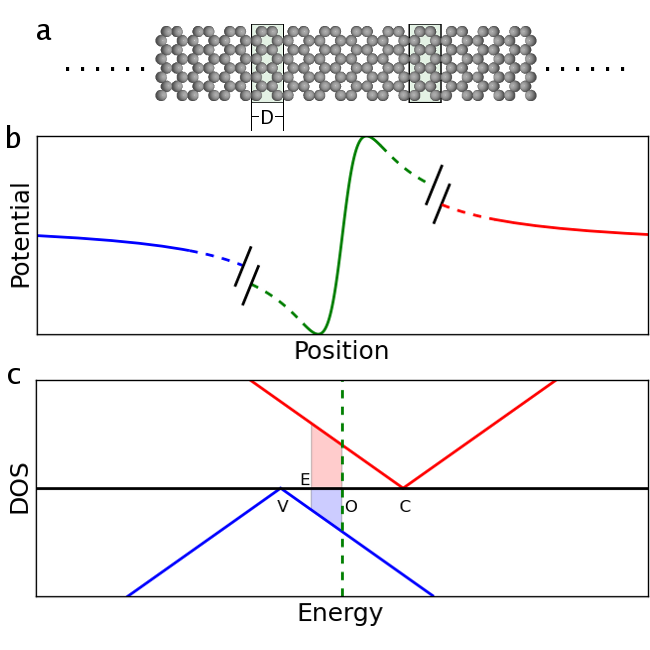}
%model_gate_vt_dos.png
\caption{(color online) (a) Gate controlled PN junction: it is a perfect graphene sheet with two local metal gates (shown as green shadowed boxes) on top. Coordinates are defined the same way as that in Fig.\ref{fig1}. (b) Schematic plot of a dipole potential along the transport (z-axis) direction. At $z=\pm\infty$, the potential equals to that of perfect graphene. Dashed lines indicate the slowly decaying potential over a long distance. (c) The DOS of gate controlled graphene PN junction. Above the horizontal black line is for a positive gate voltage (right side) where the linear Dirac bands are shifted upward; below is for a negative gate voltage (left side). The green vertical dashed line indicates the Fermi level. Shadowed areas represent the region where electrons can be excited by photons to contribute to the photocurrent.}
\label{fig4}
\end{figure}

The structure of a gate controlled device is shown in Fig.\ref{fig4}a. We imagine that a metal gate of finite width $D$ is attached on either side of the graphene forming a transport junction. When opposite gate voltages are applied on the two gates, a potential drop is established across the junction. Far left to the left-gate and far right to the right-gate, the system is not affected by the gate potentials so that the two electrodes far away still have the same chemical potential, as schematically shown in the potential profile in Fig.\ref{fig4}b. Without photons, there is no dc current because the electrochemical potentials of the left and right reservoirs are the same. With photons, electrons may be excited from the valence DOS (as indicated by the shadowed area in Fig.\ref{fig4}c) to the conduction DOS, and a net dc current is generated by the local electric field.

In practical calculations, gate voltages can be applied as the electrostatic boundary conditions for the Hartree potential\cite{taylor2}. For graphene, the potential decays very slowly due to poor screening of low dimensionality as schematically indicated in Fig.\ref{fig4}b: very long graphene sheet between the two gates would be needed in the calculation to correctly capture the potential profile. This computation difficulty can be bypassed approximately. In our analysis, we neglect electron-hole pair splitting in the decaying region of the potential, this is well justified due to the weak local electric field. We also neglect any change (due to the gate) in the self-energy: this is similar to the wide-band approximation. Hence, the gate voltages can be simulated in the NEGF-DFT method very simply as follows. We first carry out a NEGF-DFT self-consistent analysis of a homogeneous graphene, increasing the bias voltage to $U$, and then we reset the electrochemical potentials of both electrodes to the Fermi level of the unbiased graphene in subsequent photocurrent calculations. As a result, the junction produces photocurrent when the scattering region (between the gates) is under light illumination.

\begin{figure}[ht]
\centering
\includegraphics[width=0.5\textwidth]{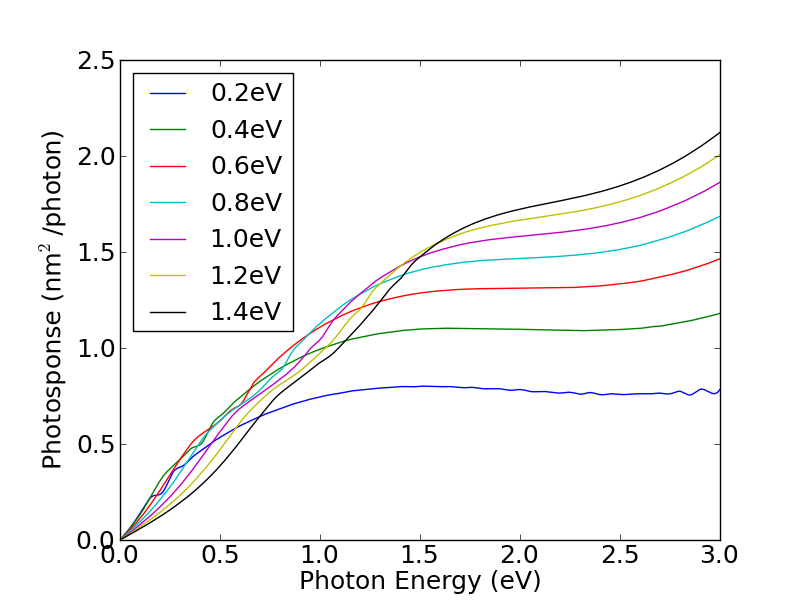}
%gate_response0.png
\caption{(color online) Photo-response function of the gate controlled graphene PN junction under different gate voltage differences $U$ displayed in the legend. On the right of this figure (higher photon energy range), the curves are ordered from low $U$ to high $U$, namely the lowest curve is $U=0.2$V and highest curve is $U=1.4$V.}
\label{fig5}
\end{figure}

In Fig.\ref{fig5} we plot the photo-response function $f$ (Eq.\ref{f1}) subjected to different gate voltage differences $U$ for the gate controlled junction, versus the photon energy $E_{ph}=\hbar\omega$. When $E_{ph} < U$, $f$ increases linearly with $E_{ph}$. In Refs.\onlinecite{Model,ModelLong}, by solving a semiclassical transport model the photo-response function $f=\frac{h\omega I}{eS}$ under certain light intensity $S$, is also found to be linear in photon energy, $f=\frac{\pi e^2W\omega}{2c\beta}\propto\omega$ when $\hbar\omega\sim U$ ($W$ is the width of the strip and $\beta$ the slope of the potential). Here we give a simple argument of the linear relationship when $E_{ph} \sim U$ - based on the intrinsic linear dispersion of graphene. As shown in Fig.\ref{fig4}c, the ``V" shaped DOS of a pristine graphene is shifted above or below the Fermi level by the gate voltage, and the distance between the two dips (labeled C and V) is the difference $U$ between the gate voltages. A photon with energy $E_{ph}$ can excite valence electrons having energies in the range [$E_{ph}$, $0$] to the conduction band, i.e., the number of excited electrons $N_{ex}$ is proportional to the shadowed area indicated in Fig.\ref{fig4}c. The excited electrons are collected by electrodes at the edge of the scattering region where the coupling Hamiltonian can be simplified as $i\Gamma$ ($\Gamma$ is a constant representing the lifetime of electron at the edge). Photocurrent is therefore determined by two factors, $I_{ph}\propto N_{ex}\Gamma$. In Fig.\ref{fig4}c, the red shadowed area means that the electron is excited from the right side to the left side of the junction, and the opposite is true for the blue shadowed area. The photocurrent contributed by these two areas have opposite sign, and the net current depends on their difference. Due to the linear dispersion of graphene, the difference of these two areas scales with $E_{ph}$, therefore the photo-response - photocurrent induced by one photon, is linear to $E_{ph}$. For $E_{ph} > U$, the response involves more complicate factors rather than the $\pi$ bands. From the results we conclude that it tends to saturate and appears quite flat over a broad range of $E_{ph}$\cite{foot2}.

\begin{figure}[ht]
\centering
\includegraphics[width=0.5\textwidth]{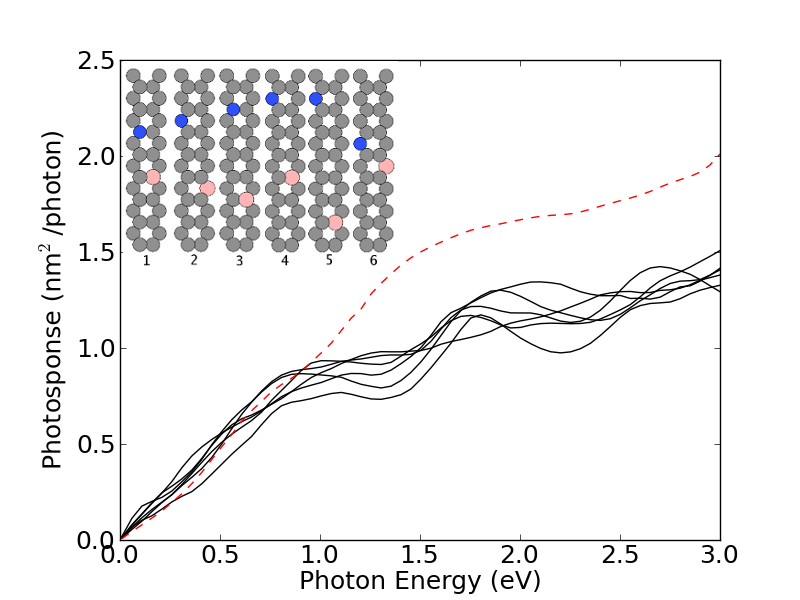}
%Locations.png
\caption{(color online) Black solid curves are the photo-response function for junctions with different dopant configuration (interface structures shown in the inset). These configurations are generated by changing the locations of P, Q sites in Fig.\ref{fig1}. The red dotted curve is the photo-response of the gate controlled PN junction having an average potential decline of 1.4eV. The polarization angle of the incident light is set to $\theta=\pi/2$ for all cases.}
\label{fig6}
\end{figure}

A very interesting issue is how does gate controlled device compare with the chemically doped ones. To this end we calculated six samples of chemically doped devices in which the dopant locations are different (positions indicated by letters P and Q in Fig.\ref{fig1}). The results of the gate controlled device and the six chemically doped devices are shown in Fig.\ref{fig6}. At low energy, the photo-responses of all the devices are similar, suggesting that the response is mostly determined by the overall potential slope of the PN junction, namely the carbon system provides the main contribution to photocurrent. When photon energy is larger than 1eV, the photo-response starts to differ from each other. The chemically doped junctions have some specific undulations (black curves in Fig.\ref{fig6}) which can be traced to dopant scattering. This is a characteristic fingerprint owing to quantum interference between different electron propagating paths. The photo-response of the gate controlled junction is smooth since the potential profile is relatively smooth\cite{foot3}. It also has a larger response in the high frequency regime. This is because in chemically doped junctions, the strong local fields near the dopants may not contribute to photocurrent if the donor and acceptor sit far apart from each other and, on average, for chemical junctions the active carbon system suffers from a weaker global field across the PN junction as compared to that of the gate controlled junctions.

\section{Discussion and Summary}

The calculated values of the photo-response function imply that for a $\sim 1mm$ wide graphene PN junction(the light is shined perpendicularly to the junction plane), at the standard solar light intensity of $0.1W/cm^2$ and a photon energy of $1eV$, the corresponding photocurrent should be in the order of $nA$ which is detectable experimentally. For the chemically doped graphene PN junctions with high dopant concentration, short-range disorder may trigger charge localization, coherent back scattering and damped quantum interference\cite{Localization_Theory}. The localization length at the Dirac point of graphene is estimated to be $\sim 200nm$ by two-dimensional scaling theory\cite{Localization}. The predicted photocurrent may be suppressed to some extent by such effects.

We mention in passing that effects of charge-transfer exciton which is important for organic PV devices has not been considered in our analysis because it is negligible for single layer graphene, especially in the low photon energy regime (< 1eV)\cite{Louie}. Finally, we did not observe quantum interference between two electron paths accompanied by resonant absorption of photons discussed theoretically in Ref.\onlinecite{QI}. The reason may lie in the mismatch of the device parameters which are required to satisfy both strong reflection and smooth potential profile, hence to observe that phenomenon it requires a device size in the order of 100nm which is much larger than the systems we investigated here.

To summarize, we have carried out a first principles atomistic analysis of the  photo-response of graphene PN junctions. For both chemically doped and gate controlled junctions, we examined several main features of the photo-response function. Our calculations show that the graphene PN junctions have a broad band photo-response - including terahertz, which is difficult to reach by conventional semiconductors. The dependence of photo-response on the angle between the direction of light polarization and the PN interface is consistent with the previous theoretical prediction\cite{Model}. The overall trend of the dependence of photo-response against photon energy, except the chemical fingerprints observed in the chemically doped junctions, agrees well with that of the semiclassical model in the small intensity regime. Importantly, the essential properties of the photo-response for the two kinds of PN junctions are found to be similar.

\section{Acknowledgment}

We gratefully acknowledge financial support from NSERC of Canada, FQRNT of Quebec and CIFAR. We thank RQCHP for providing computing facilities.


\begin{thebibliography}{32}
\bibitem{foot2}
In this work we focus on the visible spectrum of solar light and did not investigate the regime of photon energy $E_{ph} > 3.0$eV.
\bibitem{foot3}
In the numerical computation, a dense k-sampling is necessary to obtain smooth potential profiles, here 32 k-points along the x-axis were used.
\end{thebibliography}
\end{document}